\documentclass[12pt]{iopart}

\usepackage{gensymb}
\usepackage{graphicx}
\usepackage{epstopdf}
\usepackage{float}

\begin {document}

\title{Experimental Evaluation of the Claimed Coulomb Rotation (Electrostatic Torque)}
\author{D Bojiloff and M Tajmar}

\address{Institute of Aerospace Engineering, Technische Universit\"at Dresden, 01062 Dresden, Germany}
\eads{\mailto{martin.tajmar@tu-dresden.de}}

\begin{abstract}

In the year 2002 publications of A.V.M. Khachatourian and A.O. Wistrom were released, in which the existence of an electrostatic torque has been claimed. This moment of force should act in a three sphere configuration, where one sphere is held at a constant electric potential. This claim was based on an observed rotation and was supported by a mathematical solution derived by Wistrom and Khachatourian. The theoretical work of Wistrom and Khachatourian as well as the interpretation of the observed rotation were criticized by several scientists who offered alternative explanations for the rotation. We therefore designed an experimental setup which enabled us to investigate the phenomenon. By performing numerous measurements, we showed that the rotation is due to asymmetric mass distribution within the sphere, which is dislocated due to electrostatic forces between the spheres. We were able to clear our measurements from this effect and observed a null result more than two orders of magnitude smaller than predicted by Khachatourian and Wistrom's theory. We therefore showed that the rotation doesn't occur in an electrostatic system within the resolution of our experiment.

\end{abstract}


\maketitle

\section{Introduction}

\begin{figure}[H]
	\centering
	\includegraphics[angle=0, scale=0.5]{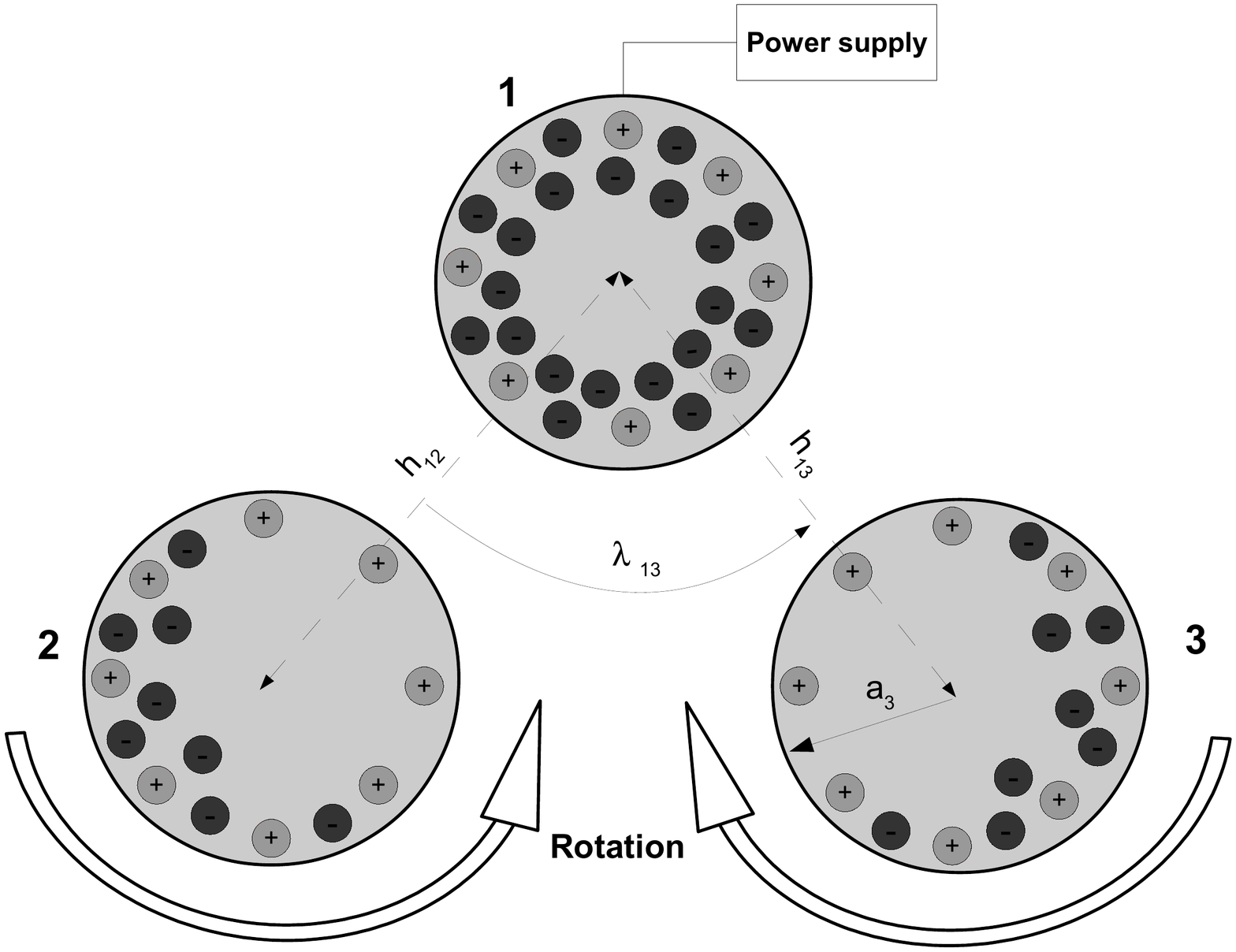}
	\caption{setup of electrostatic torque experiment (sphere one is connected to HV power supply, spheres two and three are floating)}
	\label{principle}
\end{figure}

According to Khachatourian and Wistrom, if a sphere is connected to a high potential two other spheres in a certain geometric arrangement show a torque where the two spheres rotate in a direction opposite to each other \cite{khachatourian2002, wistrom2002} as outlined in Fig.~\ref{principle}. If this effect exists, it would be a new electrostatic effect and therefore of high scientific value. Both authors also proposed a detailed theory of why such an electrostatic torque should exist similar in the order of magnitude of their observations \cite{khachatourian2000, khachatourian2003}.  As an explanation, they name electrostatic forces between charges residing on the conductor surface, which can lead to torque due to asymmetric charge distribution. In a three sphere configuration, where sphere one is put under an electric potential, the charge distribution on the others spheres will be asymmetric, as long as the spheres are not aligned linearly. This distribution arises due to repelling forces between the movable charge carriers. According to Khachatourian and Wistrom, this charge distribution causes a static moment of force acting on spheres two and three. The torque is then due to electrostatic forces acting on the charges residing on the conductor surface.

After the publication of their experimental claim and theory, immediately several authors found mathematical flaws in their derivation \cite{horoi2004, jensen2004, hense2004} rendering the predicted torque to zero. Also an alternative explanation for the observed torque was suggested by Levin \cite{levin2003} such that a low current corona discharge could produce a similar effect (Khatchatourian and Wistrom claim that current flow was negligible). However, so far no replication attempt has been made to verify the observed torque and to find a satisfactory explanation. Here we present detailed measurements of a configuration similar to the one from Khachatourian and Wistrom producing torques in the order of magnitude of their theory. After testing and evaluation of different setups, we can finally shed light on the origin of this controversially discussed effect. 

\section{Predicted Electrostatic Torques}

\begin{figure}[H]
	\centering
	\includegraphics[angle=0, scale=0.6]{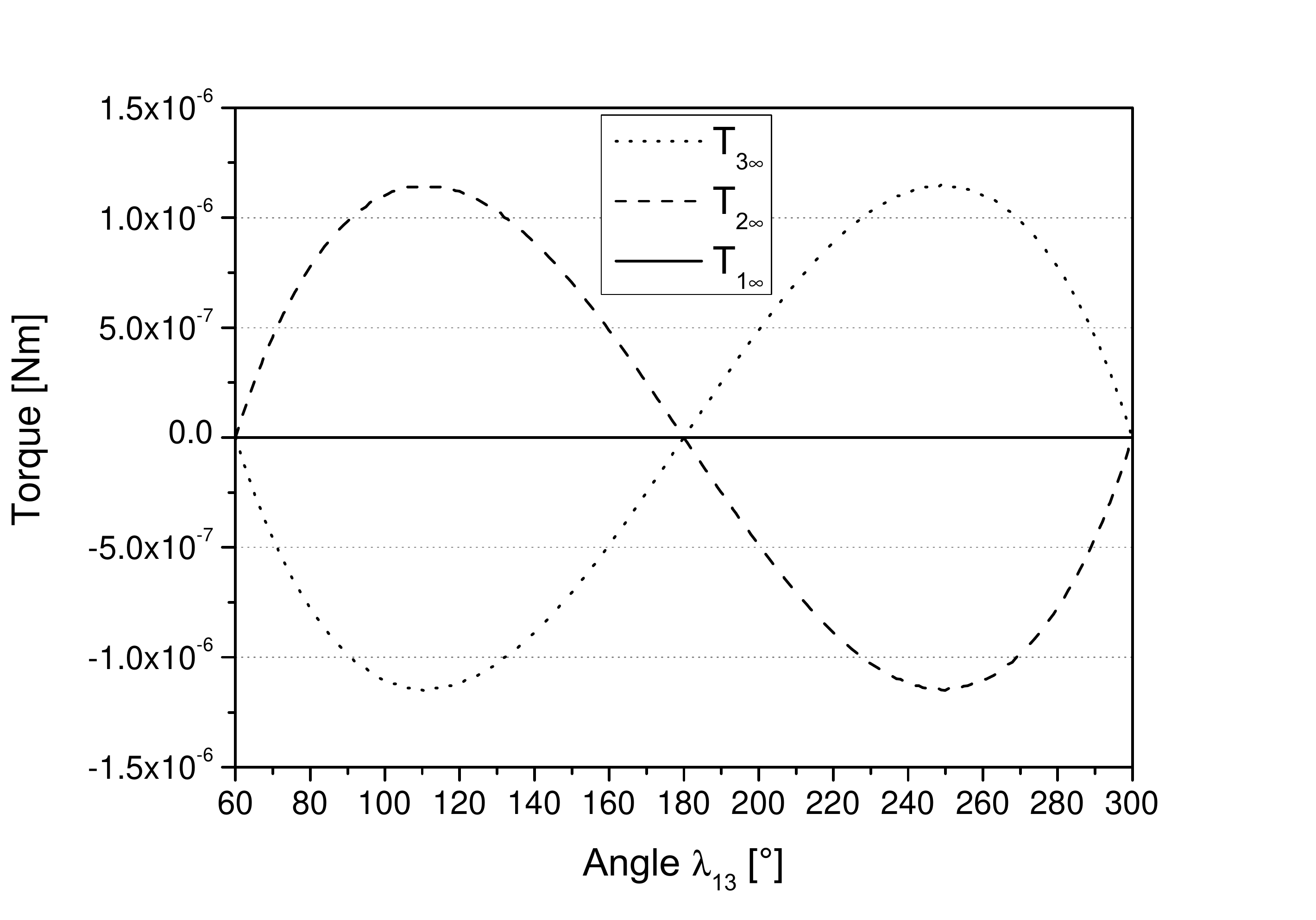}
	\caption{plot of the asymptotic value of electrostatic torque in dependence of $\lambda_{13}$ ($V_1=V_2=V_3=5kV, a=25mm, h_{13}=h_{12}=60mm$)}
	\label{coulombtorque}
\end{figure}

To calculate the electrostatic torque, Wistrom and Khachatourian used an action-at-a-distance approach. First, the charge distribution on the surface is obtained using Gauss's law and the distribution is then considered to consist of infinitesimal parts. The effect of every part on a sphere on every part of the other two spheres has then to be calculated. To do so, expressions for surface potentials are to be obtained, like described in \cite{khachatourian2000}. The coulomb torque can be obtained by integrating all effects over the whole surface. 
Using this method, Wistrom and Khachatourian \cite{khachatourian2003} obtained three expressions for the asymptotic electrostatic torque, one for each sphere, which read 

\begin{equation*}
{\bf T_{1 \infty}} = -\hat y \frac 1  K \left( \frac {A_{1,1}^1 A_{0,0}^2 - A_{0,0}^1 A_{1,1}^2} {h_{12}^2}  - \frac {A_{1,1}^1 A_{0,0}^3 - A_{0,0}^1 A_{1,1}^3} {h_{13}^2} \right) \\
\end{equation*}
\equation
{\bf T_{2 \infty}} = -\hat y \frac 1  K \left( \frac {A_{1,1}^2 A_{0,0}^3 - A_{0,0}^2 A_{1,1}^3} {h_{23}^2}  - \frac {A_{1,1}^2 A_{0,0}^1 - A_{0,0}^2 A_{1,1}^1} {h_{12}^2} \right) \\\
\endequation
\begin{equation*}
{\bf T_{3 \infty}} = -\hat y \frac 1  K \left( \frac {A_{1,1}^3 A_{0,0}^1 - A_{0,0}^3 A_{1,1}^1} {h_{13}^2}  - \frac {A_{1,1}^3 A_{0,0}^1 - A_{0,0}^3 A_{1,1}^2} {h_{23}^2} \right) ,
\end{equation*}

where $K$ is the coulomb constant and the coefficients $A^l_{n,m}$ are$^{\footnotemark[1]}$

\begin{equation*}
A^1_{0,0}= a_1 V_1 \qquad   \qquad
A^1_{1,1}= a_1 \sin \frac {\lambda_{13}} 2 \left( V_2 a_2 \frac {a_1^2} {h_{12}^2} + V_3 a_3 \frac {a_1^2} {h_{13}^2} \right) \\
\end{equation*}
\equation
A^2_{0,0}= a_2 V_2 \qquad   \qquad
A^2_{1,1}= a_2 \sin \frac {\lambda_{21}} 2 \left( V_3 a_3 \frac {a_2^2} {h_{23}^2} + V_1 a_1 \frac {a_2^2} {h_{21}^2} \right)  \\
\endequation
\begin{equation*}
A^3_{0,0}= a_3 V_3 \qquad  \qquad
A^3_{1,1}= a_3 \sin \frac {\lambda_{32}} 2 \left( V_1 a_1 \frac {a_3^2} {h_{31}^2} + V_2 a_2 \frac {a_3^2} {h_{32}^2} \right) .
\end{equation*}

where $a_i$ is the radius and $V_i$ the potential of sphere i and $h_{i,j}$ is the distance between sphere $i$ and $j$ and $\lambda_{i,j}$ is the angle between sphere i and j at sphere k. The torques are therefore dependent on the angles between the spheres. Fig.~\ref{coulombtorque} shows  the asymptotic value of the coulomb torque for spheres two and three plotted over the angle between them and sphere one.

\footnotetext[1] {We omitted the minus sign in the subscript of the coefficient A with respect to Wistrom and Khachatourian's original publication \cite{khachatourian2003} in order to be able to calculate the torque. As we get the same torque values as in their publication, we believe that the minus sign was an error.}

Experimentally, they observed a rotation of 5$^{\circ}$-10$^{\circ}$ \cite{wistrom2003} of metallic spheres with a diameter of 270 mm using steel wires with a lenght of one meter and a diameter of 127 µm and an applied potential up to 5 to 6 kV as well as a surface-to-surface distance of at least 5 mm, which can be put down to an acting torque of about $ 2 \cdot 10^{-7}$ Nm to $ 4 \cdot 10^{-7} $ Nm that is within the range defined by the calculations above. No variation with angle or separation distance was given. The asymptotic value is correct if the separation distance between the spheres is much larger than the sphere diameter - which was not the case in their measurement. Nevertheless they claimed the order of magnitude between experiment and theory. 

\section{Experimental setup}
In our approach, we decided to use smaller diameter spheres (50 mm hollow steel spheres, sanded with korn 340) with larger and variable separation distances in order to better match the theory. Moreover, the more compact setup could potentially fit into a vacuum chamber available at our lab to rule out corona discharges if necessary. The orientation of the spheres were measured with lasers and dielectric mirrors on top of the spheres. A dielectric mirror was chosen as it barely influences the electric field in comparison to a metallic mirror.

The spheres had a blind M8 thread hole. A tiny hole was drilled through the sphere, opposing to the M8 thread hole. To connect a sphere with its wire, a special clamping mechanism was used. The clamping mechanism consisted of a short threaded bar. This bar was drilled through in the longitudinal direction. A thread hole was additionally drilled through the bar, perpendicular to the first borehole. The wire could then be put into the first hole and  be fixed in place using a flat grub screw on one side and a tapered one which pushes a tiny piston forward on the other side. Using such a tapered grub screw and a piston enabled us to avoid twisting and buckling of the wire while tightening the screws to some degree. The advantage in using these bars was that the bar could be screwed into the sphere completely and therefore no disturbances of the electric field would occur due to this part. 

The sphere could be suspended within a box, which consists of two PEEK plates on the top and on the bottom, PEEK rods to increase the stability of the box and 4 acrylic side plates to shield the spheres against disturbances like turbulence. The PEEK plates featured 55 threaded holes, which could be used to suspend the spheres using clamping mechanisms at the bottom, similar to the one described above, and mechanisms of two small PEEK blocks, which could be screwed together to clamp the wire in between, on the top. One threaded hole was located directly in the center of each plate and was meant for the sphere on which high voltage was applied. The other 54 holes were located in three half circles around that center hole. This formation of threaded holes enabled us to measure the angle $\lambda_{13}$ between $0^{\circ}$ and $180^{\circ}$ in $10^{\circ}$ steps and distances to the center sphere of 50mm, 60mm and 70mm. The material PEEK for the top and the bottom plate was chosen because of its excellent dielectric properties. Clear acrylic plates were chosen for the sides to avoid wind disturbances. The bottom plate had four threaded holes in addition. These were located in the corners and allowed the usage of regular screws as adjustable legs, to align the bottom plate exactly horizontal. The tests have been made using tungsten wires with a diameter of 0.1 mm and a purity of 99.95 \%. 

\begin{figure}[H]
	\centering
	\includegraphics[angle=0, scale=0.75]{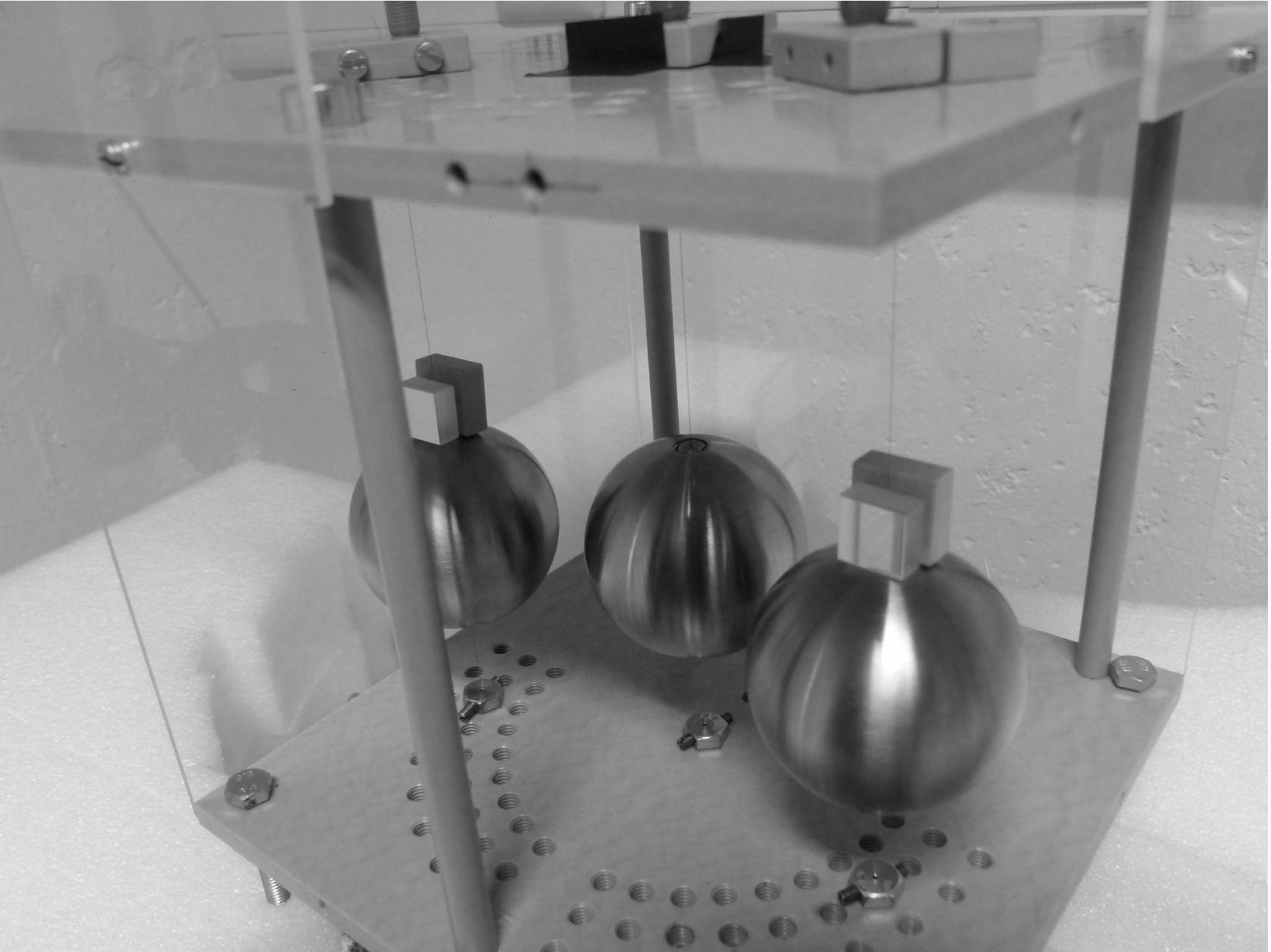}
	\caption{experimental box; two side plates opened}
	\label{experimentalbox}
\end{figure}

Quadratic dielectric mirrors with an edge length of 0.5 inch, consisting of fused silica with a dielectric constant of 3.8 were used. The mirrors, as well as counterbalancing weights made from PEEK, were glued on top of the threaded bars using a two-component dielectric glue. The electric potential could be applied to the center sphere using a pin plug connector. The jack of this connector was connected to a small copper plate between the two PEEK blocks, which were used to clamp the wire at the top. Note that sharp edges had to be avoided, since they cause peaks of the electric field strength, which may cause sparking. The clamping mechanism, as well as the suspending configuration and the experimental box are shown in Fig.~\ref{experimentalbox}.

\section{Measurements and results}

\begin{figure}[H]
	\centering
	\includegraphics[angle=0]{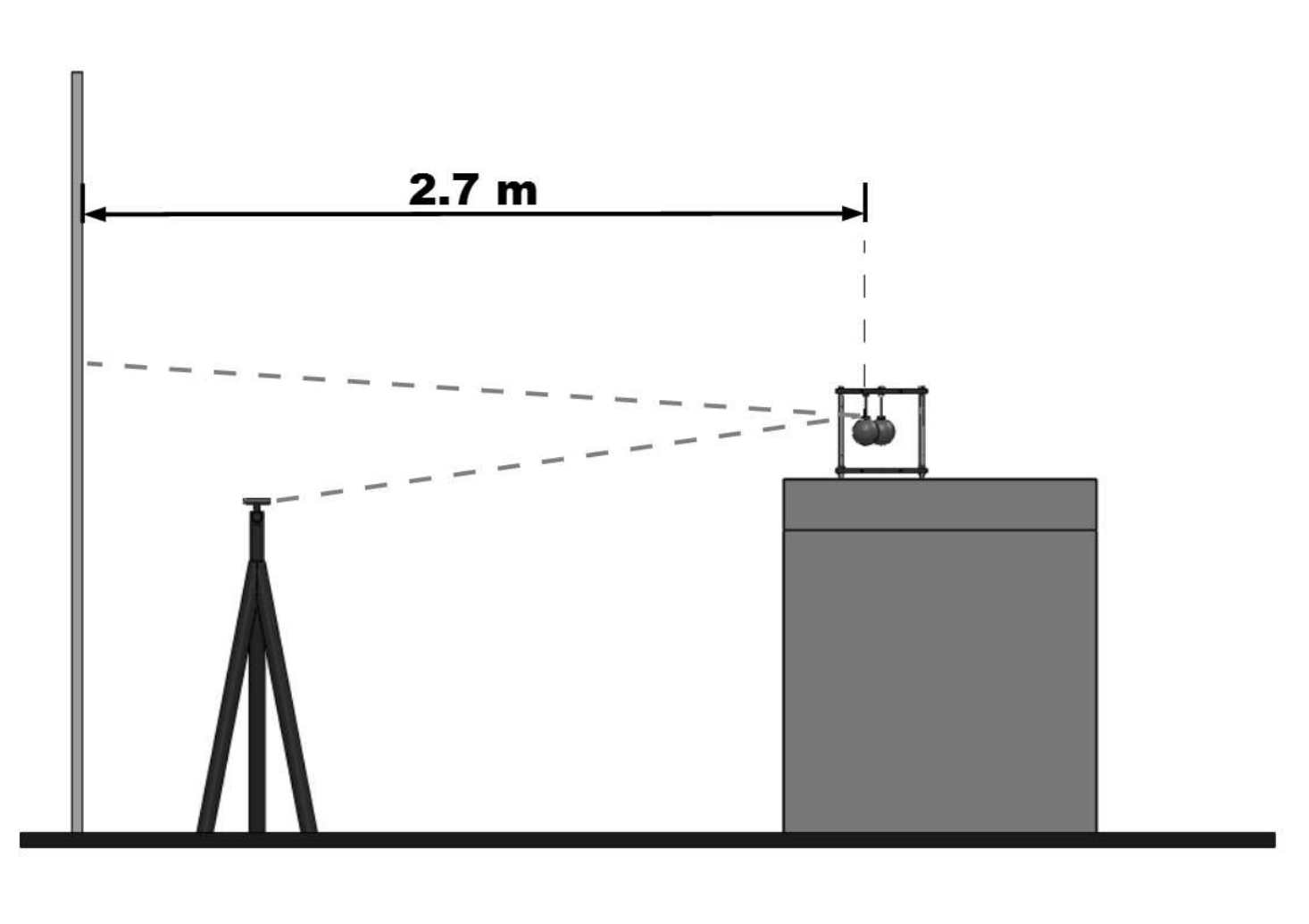}
	\caption{sketch of the experimental setup; side view}
	\label{sketch}
\end{figure}

The measurements have been done by pointing lasers onto the mirrors on top of the spheres and tracking the reflection on a screen. The distance the point traveled on the screen after an electric potential was applied was taken using a spacer and used for the calculation of the spheres rotation which yielded the acting moment of force. The distance between the box and the screen was about 2.7 m. A sketch of the experimental setup is displayed in Fig.~\ref{sketch}.

\begin{figure}[H]
	\centering
	\includegraphics[angle=0, scale=0.6]{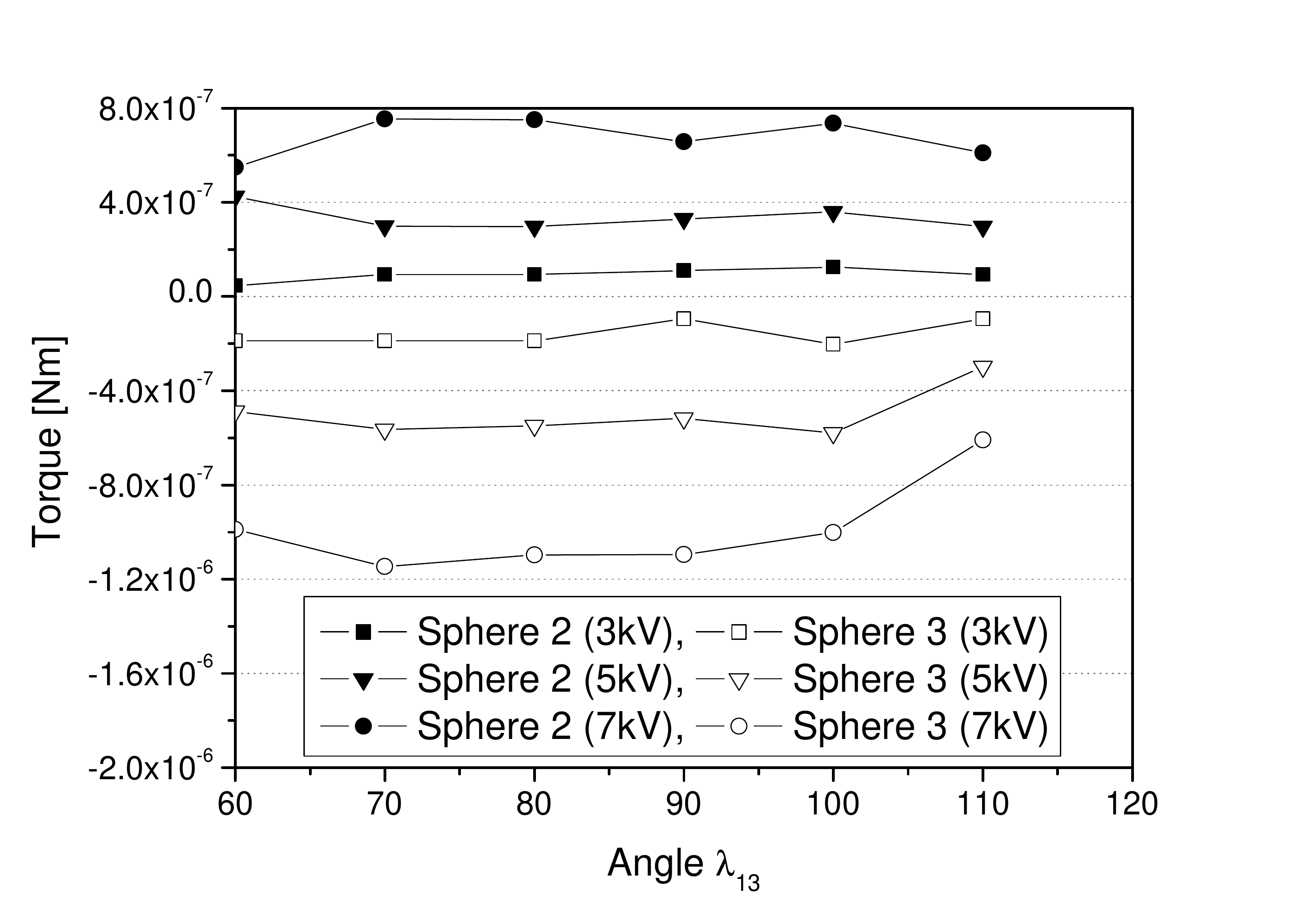}
	\caption{measured torques for different angles $\lambda_{13}$ and different voltage levels applied to sphere one, with spheres two and three floating; ($a_1=a_2=a_3=50mm, h_{12}=h_{13}=60mm$)}
	\label{angular_dependence}
\end{figure}

The first objective was to measure the torque at different angles $\lambda_{13}$ to investigate the angle-dependence of the effect. The spheres were suspended with a distance to the center sphere of h=60mm (center to center). The angle $\lambda_{13}$ was altered between $60^{\circ}$ and $110^{\circ}$ in $10^{\circ}$ steps, beginning with $60^{\circ}$. For every angle three different voltage levels were applied. Spheres two and three were floating, like in the original setup of Wistrom and Khachatourian. The resulting curve didn't show the predicted characteristic (compare Fig.~\ref{coulombtorque} and Fig.~\ref{angular_dependence}).

\begin{figure}[H]
	\centering
	\includegraphics[angle=0, scale=0.6]{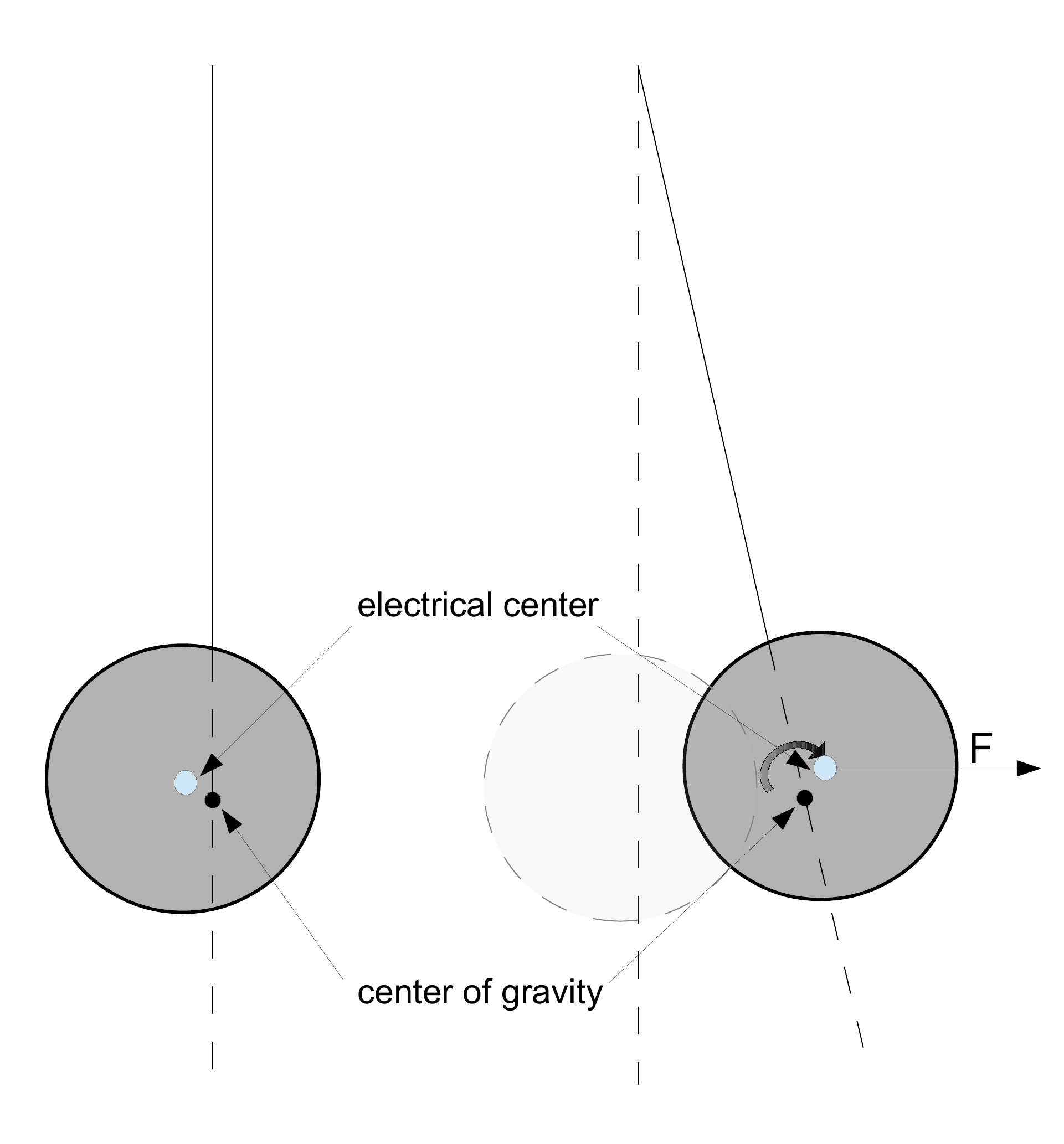}
	\caption{appearance of torque due to asymmetric mass distribution and displacement of the sphere}
	\label{asymmetric_mass_distribution}
\end{figure}

The measurements gave rise to the assumption, that the torque is due to a displacement of the sphere's center of gravity from the geometrical, hence electrical center, due to imperfect sphere geometry. Torque can occur, once a force due to electric attraction between the charged spheres acts on the electrical center of the sphere leading to a small translational movement and shifting of the rotation axis respectively the wire  (compare Fig.~\ref{asymmetric_mass_distribution}). The induced moment of force can be expressed as

\begin{equation}
{\bf  M} =  {\bf F} \times {\bf r_0}
\end{equation}

where F is the electrically induced force vector and $r_0$ is the displacement vector of the electrical center from the center of gravity.
This assumption was supported by the fact, that the absolute torque was by higher by a factor of 3 when the spheres were grounded. A grounded sphere in the vicinity of a charged one will be charged as well, since the repelled charges will leave the sphere surface through the grounding cable. The sphere as a whole is therefore charged. The attractive force between the sphere and the center sphere is then relatively high. A floating sphere as a whole is neutral, since no charges can leave the sphere surface. The attractive force between the spheres is only due to the different location of the charges on the surface and therefore relatively small. These differences regarding the force cause differences regarding the intensity of the moment of force. 

\begin{figure}[H]
	\centering
	\includegraphics[angle=0, scale=0.5]{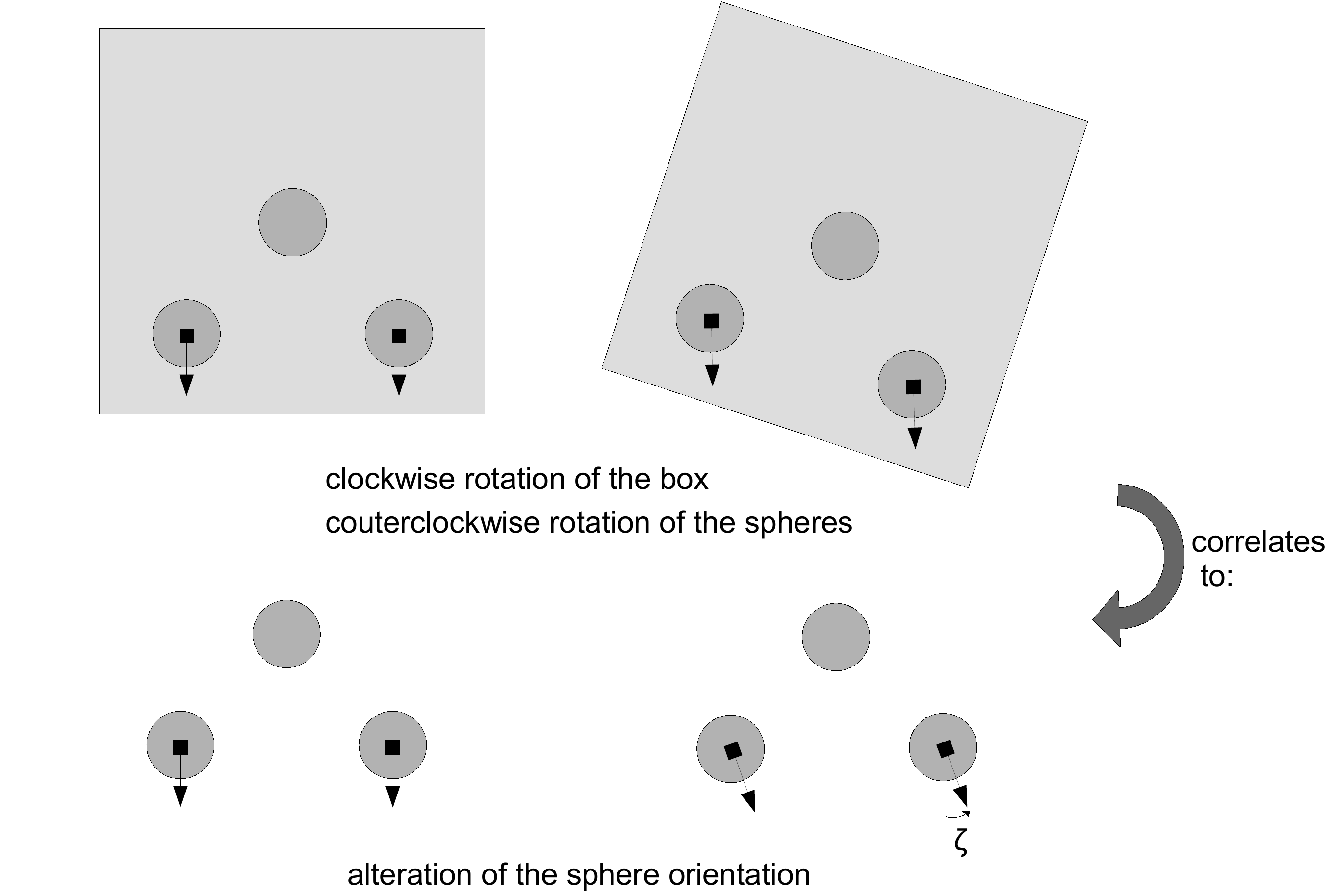}
	\caption{investigation of the influence of sphere orientation}
        \label{sphere_orientation}
\end{figure}

The measurement was aborted at $110^{\circ}$ to investigate the reliability of the measurement values and the influence of sphere orientation, and therefore of asymmetric mass distribution. To do so, the spheres were mounted four times in the same arrangement, using new wires every time. Then measurements were performed and the obtained values were compared. The experiment showed that the standard deviation was around ten percent of the average value, and therefore quite large. To test the orientation-dependence, the spheres were rotated counterclockwise in $20^{\circ}$ steps. This rotation was countered by a rotation of the whole box, such that the direction from which measurements were performed didn't change. This procedure is illustrated in  Fig.~\ref{sphere_orientation}. The rotational angle of the spheres, respectively the angle which defines the orientation of the spheres, is defined as $\zeta$ like shown below.

\begin{figure}[H]
	\centering
	\includegraphics[angle=0, scale=0.6]{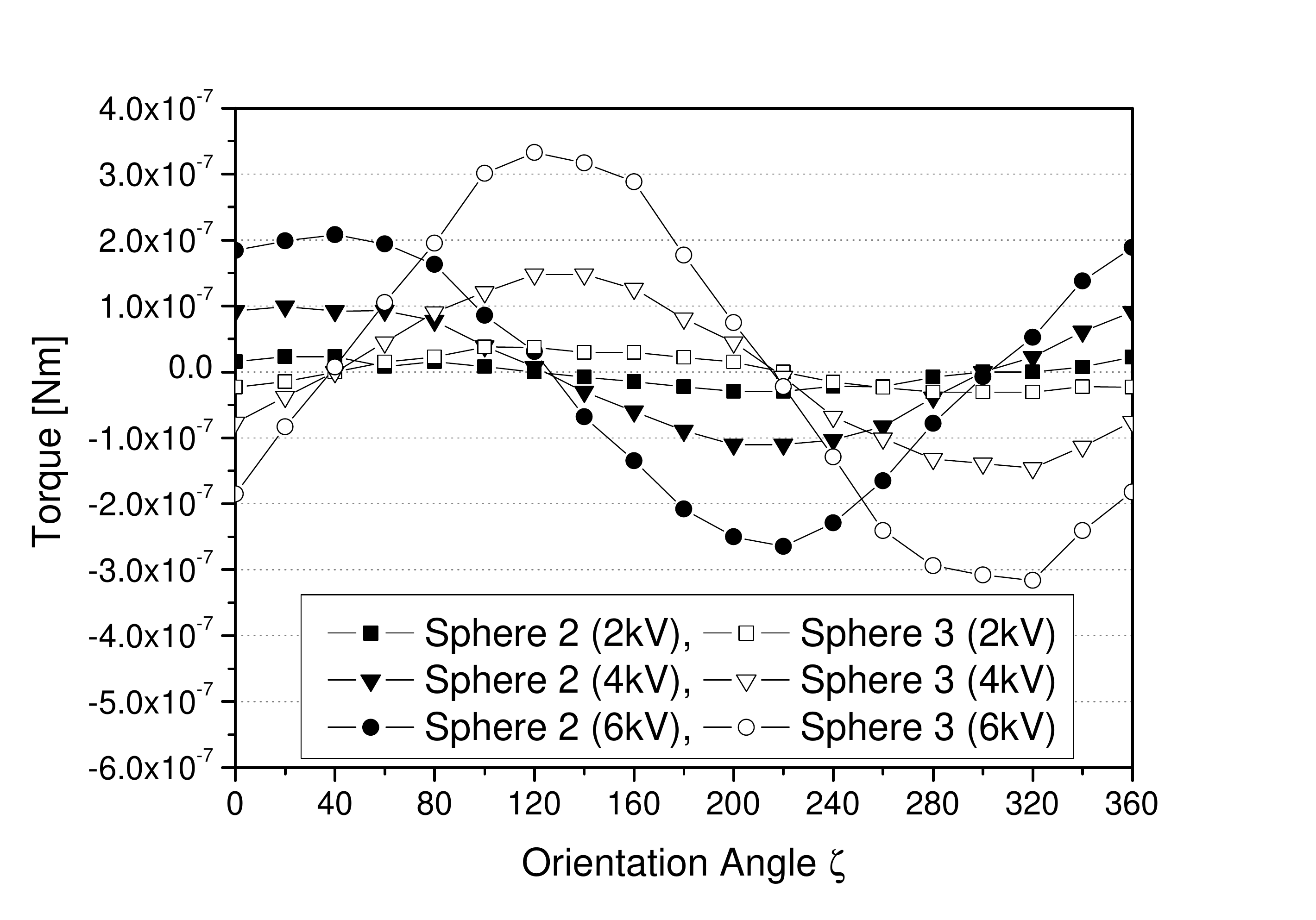}
	\caption{investigation of the orientation-dependence of the effect with 2, 4 and 8 kV applied to
	sphere one; spheres two and three floating ($a_1=a_2=a_3=50mm, h_{12}=h_{13}=60mm$)}
	\label{effect_orientation}
\end{figure}

The spheres have been rotated at the top, by simply rotating the PEEK blocks. The bottom end hasn't been rotated, to make as few alterations as possible. At each angle 2kV, 4kV and 6kV were applied. The measured values in Fig.~\ref{effect_orientation} showed, that the orientation of the spheres is of major interest. The variation of the torque over the orientation angle indicated that it was in fact the mass distribution which caused almost all of the torque. This conclusion could be obtained by the fact that the leading sign of the value changed, but the absolute value remained almost the same when the angle was altered by $180^{\circ}$.

\begin{figure}[H]
	\centering
	\includegraphics[angle=0, scale=0.6]{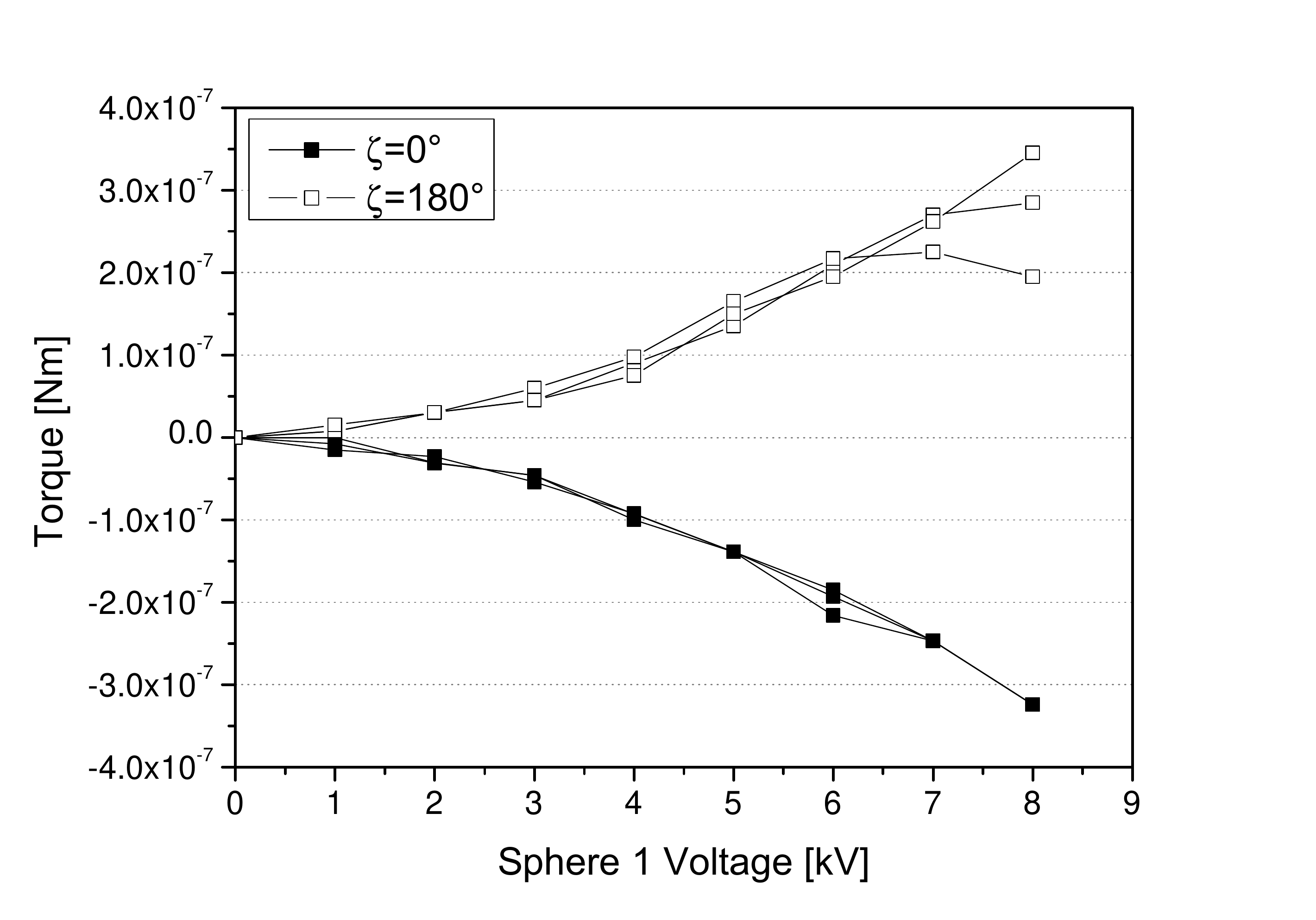}
	\caption{measured torques for sphere three at $\zeta = 0^\circ$ and $180^\circ$ with varying applied potentials to sphere one ($\lambda_{13}$ = 110$^{\circ}$), spheres two and three floating ($a_1=a_2=a_3=50mm,h_{12}=h_{13}=60mm$)}
	\label{sphere3_alternating}
\end{figure}

To eliminate the influence of the asymmetric mass distribution, it would be best to perform multiple measurements with a varying sphere orientation between 0$^{\circ}$ and 360$^{\circ}$ in very small steps. By averaging these values, a value could be obtained, which is free from this error source. Such an extensive test series was unfortunately not possible with the existing setup, since it hasn't been designed for such an application. Instead, multiple measurements were performed with two different orientation angles, $\zeta$ = 0$^{\circ}$ and $\zeta$ = 180$^{\circ}$. By turning the sphere 180$^{\circ}$ around its own axis and averaging the two values, the influence of the asymmetric mass distribution should vanish as well. By performing multiple tests and averaging the values, a more reliable value and an associated confidence interval could be obtained. Therefore 12 measurements, six for each orientation were performed for each investigated configuration. To test Wistroms and Khachatourians work, four different configurations were tested using this method. Again, the spheres were suspended in a distance of 60mm to the center sphere (center to center). The angle  $\lambda_{13}$ was set to 60$^{\circ}$, 80$^{\circ}$, 110$^{\circ}$ and 140$^{\circ}$. The voltage was increased from 0 to 8~kV in 1~kV steps. To adjust to zero, the voltage was set back to 0V after every step and it was made sure that the spheres start the next measurement without excess charge. Spheres two and three were floating throughout the measurements. Fig.~\ref{sphere3_alternating} shows the measured torques for sphere three in alternating orientations for varying applied potential to sphere one. It is noticeable that the measured torque dropped between 6 and 8 kV for some measurements, but not for all of them. This might have been due to a corona discharge and some geometrical differences between the configurations.

\begin{figure}[H]
	\centering
	\includegraphics[angle=0, scale=0.6]{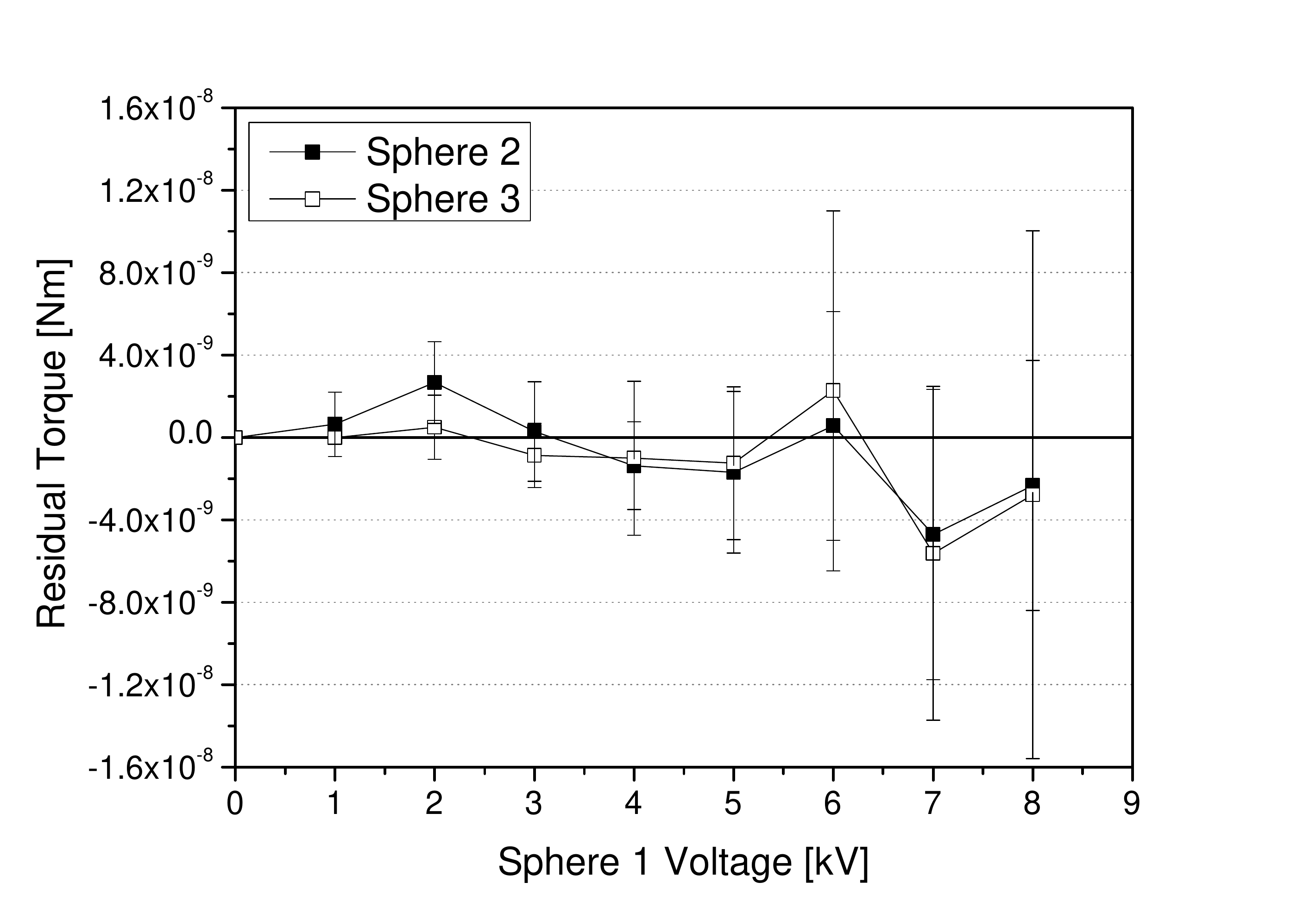}
	\caption{residual torques for $\lambda_{13}$ = 140$^{\circ}$ with varying potential applied to sphere one; spheres two and three floating ($a_1=a_2=a_3=50mm, h_{12}=h_{13}=60mm$)}
	\label{residualtorque}
\end{figure}

Fig.~\ref{residualtorque} shows the resulting torques for one of the four tested configurations ($\lambda_{13}$ = 140$^{\circ}$). These torques were obtained by averaging 12 measurements (6 at $\zeta$ = 0$^{\circ}$ and 6 at $\zeta$ = 180$^{\circ}$) in each case and are therefore free of the influence of asymmetric mass distribution. The corrected torque will henceforth be called residual torque.

\begin{figure}[H]
	\centering
	\includegraphics[angle=0, scale=0.6]{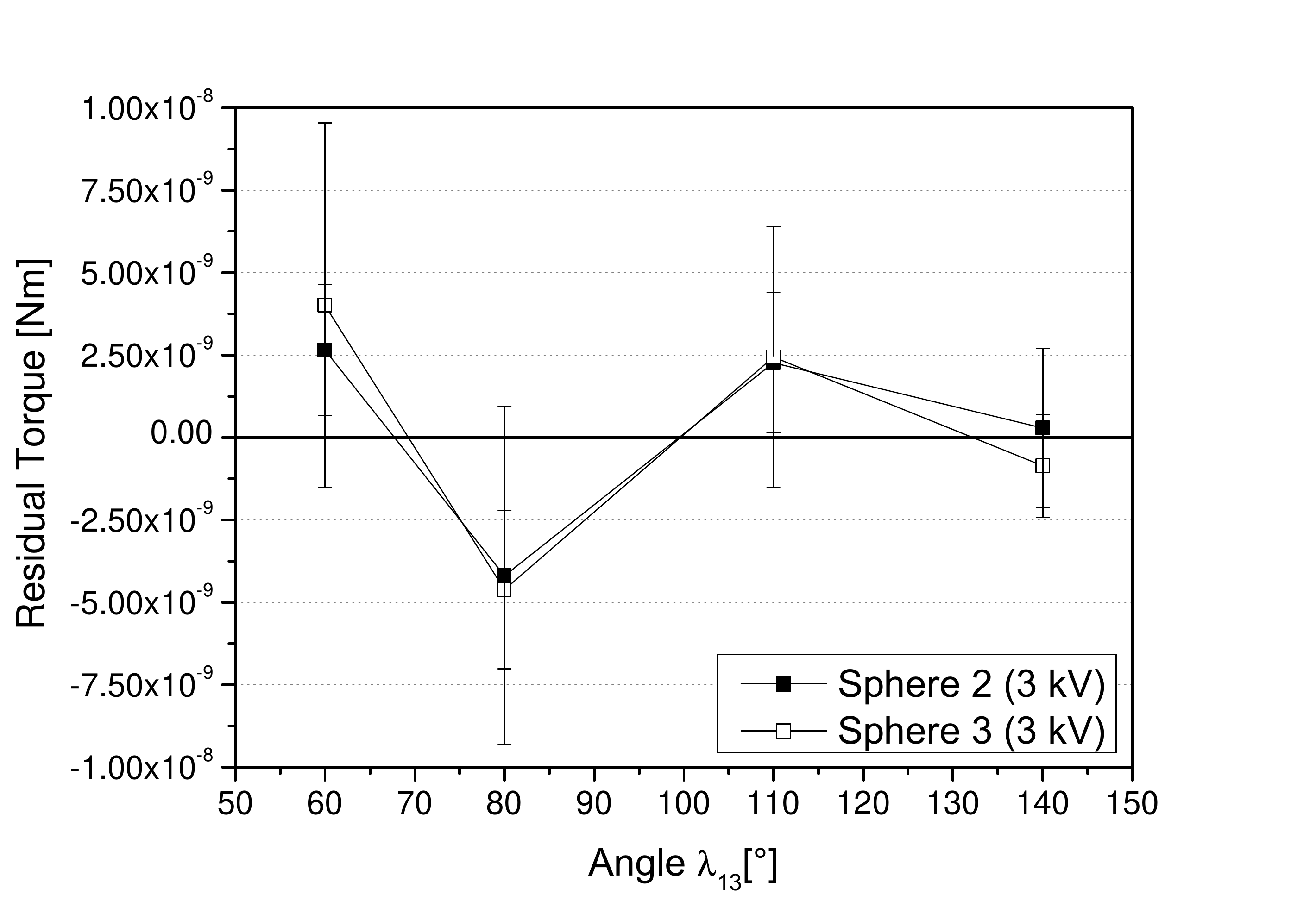}
	\caption{residual torques for different angle $\lambda_{13}$ with 3 kV applied to sphere one; spheres two and three floating ($a_1=a_2=a_3=50mm, h_{12}=h_{13}=60mm$)}
	\label{residualtorque_3kV}
\end{figure}

\begin{figure}[H]
	\centering
	\includegraphics[angle=0, scale=0.6]{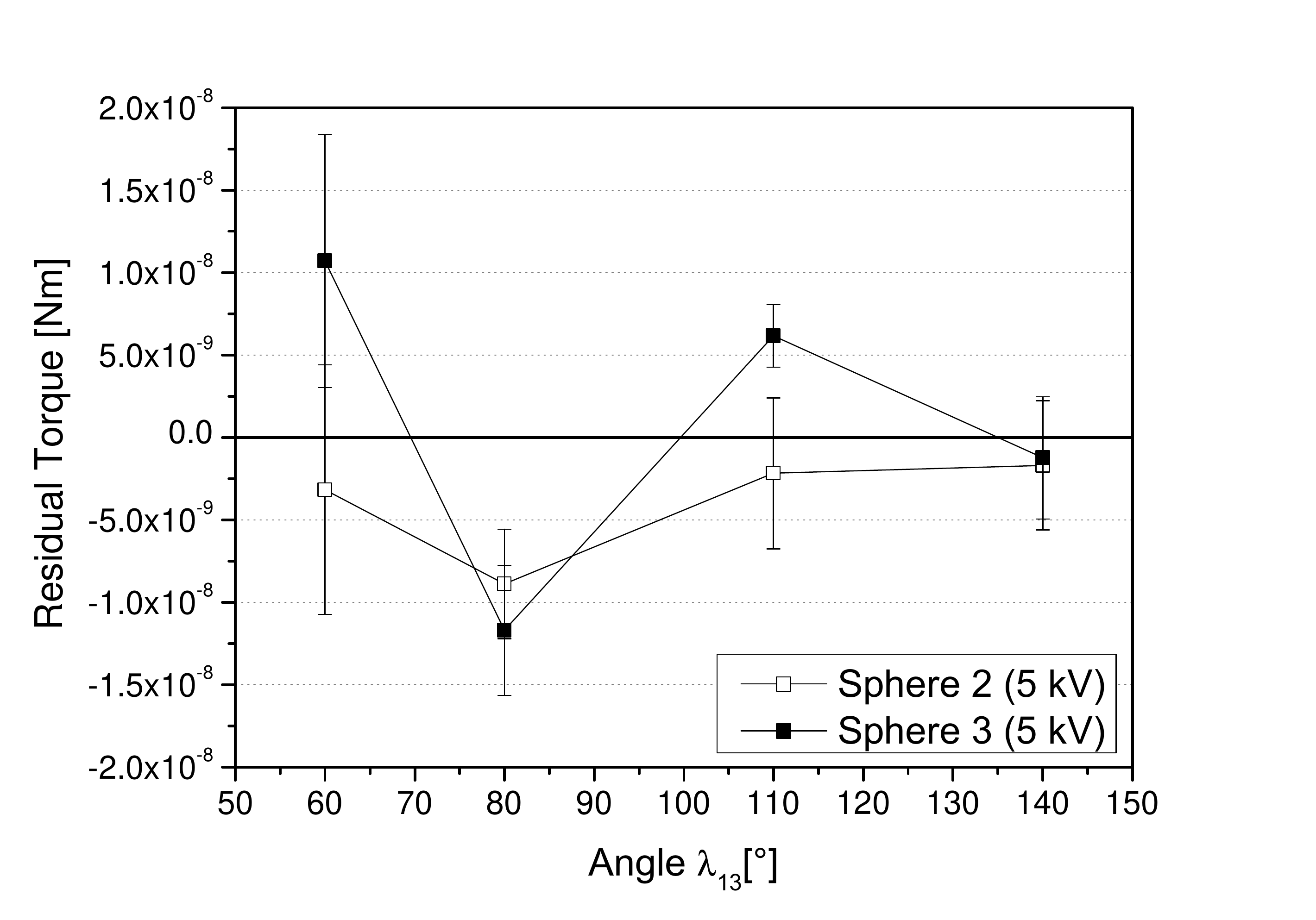}
	\caption{residual torques for different angle $\lambda_{13}$ with 5 kV applied to sphere one; spheres two and three floating ($a_1=a_2=a_3=50mm, h_{12}=h_{13}=60mm$)}
	\label{residualtorque_5kV}
\end{figure}

At $\lambda_{13}$ = 140$^{\circ}$, the measured absolute torques were very small ($\textless 5 \cdot 10^{-9}$ Nm). Considering the error bars, no residual torque is actually left within our measurement resolution which is two orders of magnitude smaller than the experimental values claimed by Khachatourian and Wistrom. The curves for sphere two and three are very similar in this case. We then chose two voltage levels (3 and 5 kV) and varied the angle $\lambda_{13}$. The residual torque analysis for both potentials is shown in Figs.~\ref{residualtorque_3kV} and \ref{residualtorque_5kV} respectively. Also here, no remaining residual torque could be found within 3$\sigma$ error bars. 

\begin{figure}[H]
	\centering
	\includegraphics[angle=0, scale=0.6]{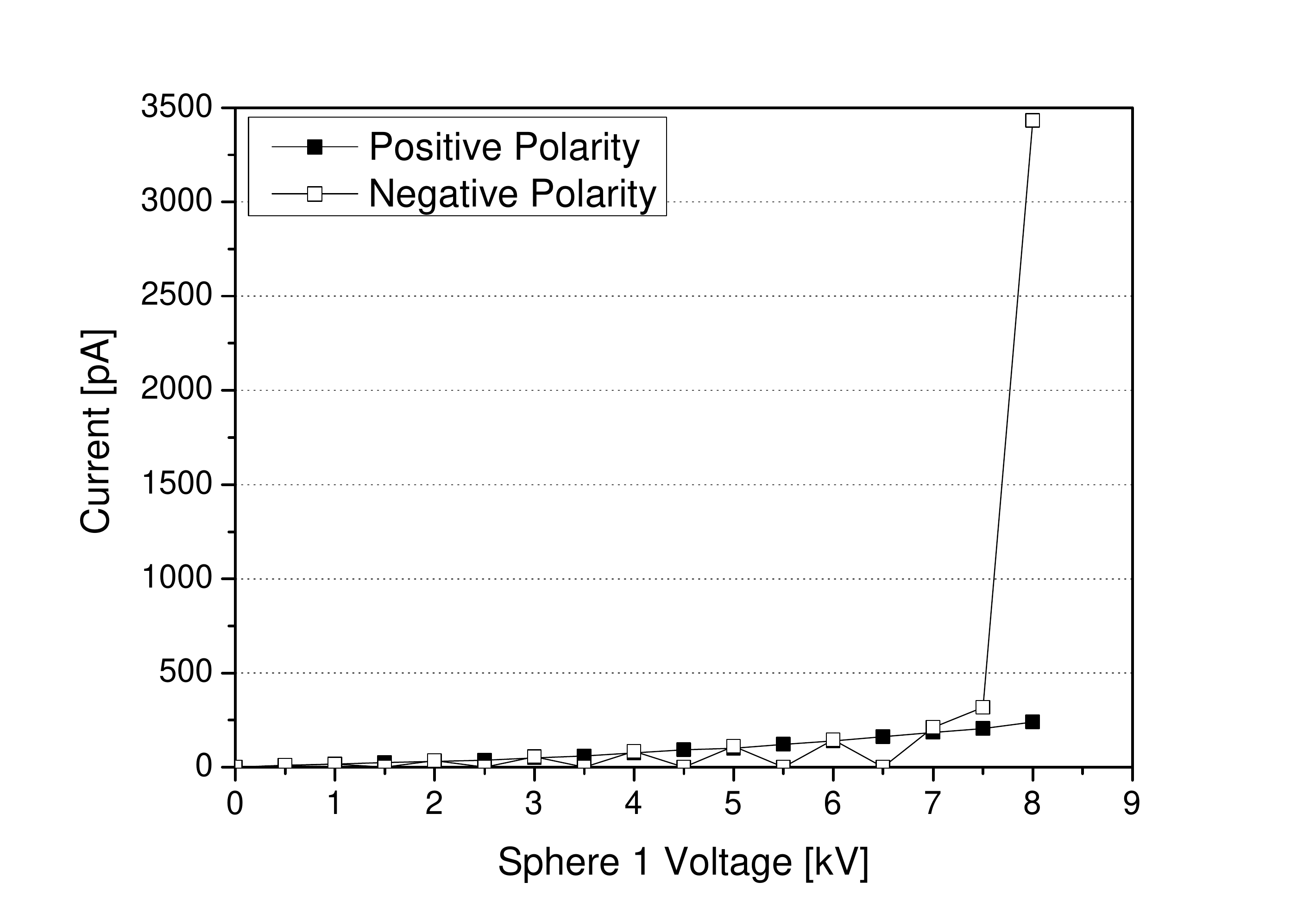}
	\caption{current-voltage characteristic; varying potential applied to sphere one and $\lambda_{13}$ = 110$^{\circ}$; spheres two and
three grounded ($a_1=a_2=a_3=50mm, h_{12}=h_{13}=60mm$)}
	\label{currents}
\end{figure}

To investigate the possibility of a corona discharge, a Keithley 6487 pico-amperemeter was used. Here we grounded sphere two and three via the amperemeter to ground (different to the floating measurements). The currents are expected to be larger than in the floating configuration as outlined above, so the obtained values can serve as an upper-limit of possible corona currents. Positive as well as negative voltage were applied and the measured currents are shown in Fig.~\ref{currents}. 

Up to about 7 kV, practically no current was measured ($\leq$ 250 pA, creeping currents over the isolators). Only in the case of negative polarity, a sharp rise was noticed after 7 kV into the nA range. This may explain the small anomalies noted in the residual torque measurements at 8 kV  in Fig.~\ref{residualtorque}. However, this test shows that corona currents are not responsible for the claimed phenomenon as suggested by Levin \cite{levin2003}. 

\section{Conclusion}

Our measurements showed that there is no residual torque within the limits of accuracy of the measurements ($\textless 3 \sigma$). The comparison of the results with the claims of Wistrom and Khachatourian showed that the measured values were smaller by two to three magnitudes (compared at 5 kV). The moment of force in the region of 10$^{-7}$ Nm, which has been calculated based on Wistrom's observations, could be reproduced in our experiment under similar conditions. This momentum could be traced back to the influence of asymmetric mass distribution within the sphere. Our measurements with the pico-amperemeter showed that there was a small current, but a corona discharge as source of the residual torque could nevertheless be ruled out. 

\section{References}

\bibliographystyle{unsrt}

\bibliography{Bibliography}

\end{document}